\documentstyle[12pt]{article}

\def\beqa{\begin{eqnarray}}
\def\eeqa{\end{eqnarray}}
\def\beq{\begin{equation}}
\def\eeq{\end{equation}}

\def\pr{{\it Phys. Rev.}\ }
\def\prl{{\it Phys. Rev. Lett.}\ }
\def\pl{{\it Phys. Lett.}\ }
\def\np{{\it Nucl. Phys.}\ }

\def\ijmp{{\it Int. Journ. Mod. Phys.}\ }

\def\grg{{\it Gen. Relativ. Grav.}\ }

\def\apj{{\it Ap. J.}\ }

\def\araa{{\it Ann. Rev. Astr. Ap.}\ }

\def\etal{{\it et al.}}
\def\ie{{\it i.e. }}
\def\eg{{\it e.g. }}

\def\p{\phi}

\def\v{V(\phi)}

\def\l{\cal L}

\begin{document}

\begin{titlepage}
        \title{Newtonian limit of String-Dilaton Gravity}
\author{S. Capozziello\thanks{E-mail: capozziello@sa.infn.it},~~
and G. Lambiase\thanks{E-mail: lambiase@sa.infn.it} \\
{\em Dipartimento di Scienze Fisiche "E.R. Caianiello"} \\
 {\em Universit\'a di Salerno, 84081 Baronissi (Sa), Italy.} \\
 {\em Istituto Nazionale di Fisica Nucleare, Sez. di Napoli, Italy.} \\ }
\date{\today}
\maketitle

\begin{abstract}
We study the weak-field limit of string-dilaton gravity and
derive corrections to the Newtonian potential which strength
directly depends on the self interaction potential and the
nonminimal coupling of the dilaton scalar field. We discuss also
possible astrophysical applications of the results, in particular
the flat rotation curves of spiral galaxies.

\end{abstract}
 \vspace{10.mm}
 PACS number(s): 04.50.+h, 04.20. Cv, 98.80. Hw \\
 \vspace{5.mm}
 Keyword(s): string--dilaton theory,  Newtonian limit.
 \vfill
 \end{titlepage}

\section{\normalsize\bf Introduction}
String-dilaton gravity seems to yield one of he most promising
scenarios in order to solve several shortcomings of standard and
inflationary cosmology \cite{vafa}. First of all, it addresses
the problem of initial singularity which is elegantly solved by
invoking a maximal spacetime curvature directly related to the
string size \cite{gasperini}.

Besides, it introduces a wide family of cosmological solutions
which comes out thanks to the existence of the peculiar symmetry
called {\it duality} which holds at string fundamental scales as
well as at cosmological scales \cite{veneziano}. In practice, if
$a(t)$ is a cosmological solution of a string--dilaton model,
also $a^{-1}(t)$ has to be one by a time reversal $t\rightarrow
-t$. In this case, one can study the evolution of the universe
towards  $t\rightarrow+\infty$ as well as towards
$t\rightarrow-\infty$. The junction of these two classes of
solutions at some maximal value of curvature  (considered as
branches of more general solutions)  eventually should be in
agreement with inflationary paradigm and solve the initial value
and singularity problems of standard cosmological model
\cite{gasperini}.

The main interest for string-dilaton cosmological models is
related to the fact that they come from the low--energy limit of
(super)string theory which can be considered  one of the most
serious attempt, in the last thirty years, to get the great
unification. This theory avoids the shortcomings of quantum field
theories due, essentially, to the point--like nature of particles
(renormalization) and includes gravity in the same conceptual
scheme of the other fundamental interactions (the graviton is
just a string {\it mode} as the other gauge bosons
\cite{schwarz}).

However, despite of theoretical results, we are very far from the
possibility to test experimentally the full predictions of the
theory. The main reason for this failure is that the Planck
scales, where the string effects become relevant, are too far
from the experimental capabilities of today high--energy physics.
Cosmology remains the only open way to observational
investigations since the today detectable {\it remnants} of
primordial processes could be a test for  the theory.
Furthermore, a lot of open questions of astrophysics, as dark
matter, relic gravitational wave background, large-scale
structure, primordial magnetic fields and so on could be solved
by strings and their dynamics (see, for example \cite{gasperini}
and references therein).

The key element of string--dilaton gravity, in low--energy limit,
is the fact that a dynamics, consistent with duality, can be
implemented only by taking into account massless modes (zero
modes) where the scalar mode (the dilaton) is nonminimally
coupled to the other fields. The tree--level action, in general,
contains a second--rank symmetric tensor field (the metric), a
scalar field (the dilaton) and a second--rank antisymmetric
tensor field (the so--called Kalb--Ramond {\it universal} axion).
Such an action can be recast as a scalar--tensor theory, \eg
induced gravity, where the gravitational coupling is a function
of the dilaton field \cite{vafa,capozziello1}. Then it is
legitimate to study the Newtonian limit of the string--dilaton
gravity to see what is its behaviour in the weak--field and
slow--motion approximations. This approach is useful if we want to
investigate how string--dilaton dynamics could affect shorter
scales than cosmological ones.  The issue is to search for
effects of the coupling and the self--interaction potential of
dilaton also at scales of the order of Solar System or Galactic
size.

This fact is matter of debate since several relativistic theories {\it do
not} reproduce General Relativity's results but generalize them introducing
corrections to the Newtonian potential which could have interesting physical
consequences. For example, some theories give rise to terms capable of
explaining the flat rotation curve of galaxies without using dark matter as
the fourth--order conformal theory proposed by Mannheim \etal
\cite{mannheim}. Others use Yukawa corrections to the Newton potential for
the same purpose \cite{sanders}.

Besides, indications of an apparent, anomalous, long--range acceleration
revealed from the data analysis of Pioneer 10/11, Galileo, and Ulysses
spacecrafts could be framed in a general theoretical scheme by taking
into account Yukawa--like or higher order corrections to the Newtonian
potential \cite{anderson}.

In general, any relativistic theory of gravitation can yield
corrections to the Newton potential  which, in the post-Newtonian
(PPN) formalism, could furnish tests for the same theory
\cite{will}.

In this paper, we want to discuss the Newtonian limit of
string--dilaton gravity. We develop our calculations in the string
frame since we want to see what is the role of dilaton--nonminimal
coupling in the recovering of Newtonian limit.

In Sec.2, we write down the string--dilaton field equations. The
weak field approximation and the resolution of linearized
equations are given in Sec.3. In Sec.4, we discuss the results
specifying the possible astrophysical applications.

\section{\normalsize\bf The string--dilaton field equations}

The tree--level string--dilaton effective action, \ie at the lowest order in
loop expansion, containing all the massless modes, without higher--order
curvature corrections of order $\alpha'$ (\ie without the Gauss--Bonnet
invariant) is

 \begin{eqnarray}
 {\cal A}&=&-\frac{1}{2\lambda_s^{d-1}}\int d^{d+1}x \sqrt{-g}e^{-\phi}\left[R+
 (\nabla\phi)^2 -
 \frac{1}{12}H_{\mu\nu\alpha}H^{\mu\nu\alpha}+
 V(\phi)\right]+ \nonumber \\
  & &+\int d^{d+1}x \sqrt{-g}{\cal L}_m \label{1}\,,
 \end{eqnarray}
where $R$ is the Ricci scalar,  $\phi$ is the dilaton field,
$\v$ the dilaton self--interaction potential.
$H_{\mu\nu\alpha}=\partial_{[\mu}B_{\nu\alpha]}$
is the full antisymmetric derivative of the Kalb--Ramond axion tensor,
 ${\l}_{m}$ is the
 Lagrangian density of other generic matter sources.
The theory is formulated in $d+1$--dimensions and $\lambda_s$ is
the string fundamental minimal length parameter. The effective
gravitational coupling, to lowest order, is given by

\beq \label{2} \sqrt{8\pi
G_{N}}=\lambda_{p}=\lambda_{s}e^{\phi/2}\,, \eeq where $G_N$ is
the Newton constant and $\lambda_{p}$ is the Planck length. We
choose units such that $2\lambda_{s}^{d-1}=1$ so that $\exp\phi$
is the $(d+1)$--dimensional gravitational coupling. At the end of
Sec.3, we will restore standard units.

The field equations are derived by varying the action (\ref{1}) with respect
to $g_{\mu\nu}$, $\phi$, and $B_{\mu\nu}$. We get, respectively,

 \begin{equation}\label{3}
 G_{\mu\nu} +\nabla_{\mu}\nabla_{\nu}\phi+\frac{1}{2}g_{\mu\nu}\left[
 (\nabla\phi)^2-2\nabla^2\phi-V(\phi)+\frac{1}{12}H_{\alpha\beta\gamma}H^{\alpha\beta\gamma}
 \right]-\frac{1}{4}H_{\mu\alpha\beta}H_{\nu}^{\phantom{\nu}\alpha\beta}=
 \end{equation}
 $$
 =\frac{e^{\phi}}{2}T_{\mu\nu},,
 $$
 \begin{equation}\label{4}
R+2\nabla^2\phi-
 (\nabla \phi)^2 +V-V'
-\frac{1}{12}H_{\mu\nu\alpha}H^{\mu\nu\alpha}=0\,,
 \end{equation}
\begin{equation}\label{5}
  \nabla_{\mu}\left(e^{-\phi}H^{\mu\alpha\beta}\right)=0\,,
\end{equation}
where ${\displaystyle
G_{\mu\nu}=R_{\mu\nu}-\frac{1}{2}\,g_{\mu\nu} R}$ is the Einstein
tensor, $T^{\mu\nu}$ is the  stress--energy tensor of matter
sources and $V'=dV/d\phi$.

We are assuming that standard matter is a perfect fluid minimally
coupled to the dilaton. Otherwise, in the nonminimally coupled
case, we should take into account a further term in Eq.(\ref{4}).

The above ones are a system of tensor equations in $d+1$ dimensions which
assigns the dynamics of $g_{\mu\nu}$, $\phi$, and $B_{\mu\nu}$. Now we take
into account the weak field approximation in order to derive the PPN limit
of the theory.

\section{\normalsize\bf The weak field limit  and the solution of linearized
equations}
 As it is obvious, all the invariances of the full theory are not preserved
if we linearize it. For example, we loose duality in the linearized solutions.
However, this is not a problem in the present context since we are assuming a
regime well far from early singularity where duality is adopted  to solve
cosmological shortcomings. Here, we want to investigate if {\it remnants} of
primordial string--dilaton dynamics are detectable at our nearest scales
(Solar System or Galaxy).

To recover the Newtonian limit, we write the metric tensor as
 \begin{equation}\label{6}
 g_{\mu\nu}=\eta_{\mu\nu}+h_{\mu\nu}\,,
 \end{equation}
 where $\eta_{\mu\nu}$ is the Minkowski metric and
$h_{\mu\nu}$ is a small correction to it.
In the same way, we
define the scalar field $\psi$ as a perturbation of the original
field $\p$, that is
 \begin{equation}\label{7}
 \phi=\phi_0+\psi\,,
 \end{equation}
where $\phi_{0}$ is a constant. This assumption means that at
scales where Newtonian limit holds, the effects of dilaton are
small perturbations. However, as $\psi\rightarrow 0$, the
standard gravitational coupling of General Relativity has to be
restored. For the scalar potential, we can assume a power law
expression of the form ${\displaystyle \v=\lambda\p^n }$ so that,
in the same limit of (\ref{7}), we have
 \begin{equation}\label{8}
 V(\phi)\simeq\lambda \left(\phi_0^{n}+n\,\phi_0^{n-1}\psi+
 \frac{n(n-1)}{2}\,\phi_0^{n-2}\psi^2\right)\,.
 \end{equation}
 At this point,  it is worthwhile to note that the parameters
  $\lambda$ and $n$ can be related to the number of spatial
  dimension $d$ as it is shown in \cite{scud} for scalar-tensor
  theories of gravity. Below, we will show that suitable choices of
  $\lambda$ and $n$ give rise to interesting physical behaviours
  for the gravitational potential.

Finally, the weak field approximation for the axion  gives only
second order terms in the field equations with respect to
$h_{\mu\nu}$ and $\psi$ so that we can discard its contribution
in the following considerations. A physical interpretation of
this fact could be that the production of primordial magnetic
fields, considered as ''seeds" for the today observed large
magnetic fields of galaxies \cite{giovannini} is a second order
effect if due to $H_{\mu\nu\alpha}$. This topic will be studied
elsewhere.

Let us define now the auxiliary fields

 \begin{equation}\label{9}
 \overline{h}_{\mu\nu}\equiv h_{\mu\nu}-\frac{1}{2}\,\eta_{\mu\nu} h\,,
 \end{equation}
 and
 \begin{equation}\label{10}
 \sigma_{\alpha}\equiv {\overline h}_{\alpha\beta,\gamma}
 \eta^{\beta\gamma}\,.
 \end{equation}
The Einstein tensor $G_{\mu\nu}$ becomes, at first order in
$h_{\mu\nu}$,

 \begin{equation}\label{11}
 G_{\mu\nu}\simeq -\frac{1}{2} \left[\Box_{\eta} {\overline
 h}_{\mu\nu}+\eta_{\mu\nu}\sigma_{\alpha,}^{\phantom{,\alpha}\alpha}-
 \sigma_{\mu,\nu}-\sigma_{\nu,\mu}\right]\,,
 \end{equation}
 where $\Box_{\eta}\equiv \eta^{\mu\nu}\partial_{\mu}
\partial_{\nu}$. We have not fixed the gauge yet.

Using the approximation  (\ref{7}), and the approximated
expression of the scalar potential (\ref{8}),
 the field equations (\ref{3}) and (\ref{4}) become, respectively,
\beqa
  \Box_{\eta}{\overline h}_{\mu\nu}+\eta_{\mu\nu}\sigma_{\alpha,}^{\phantom{,\alpha}\alpha}-
 (\sigma_{\mu,\nu}+\sigma_{\nu,\mu})&-&2\psi_{,\,\mu\nu}+2\eta_{\mu\nu}
 \Box_{\eta}\psi+\\
 \nonumber
 &+&\lambda(\eta_{\mu\nu}+h_{\mu\nu})\phi_0^n+\lambda n\phi_0^{n-1}\eta_{\mu\nu}\psi\simeq
  -e^{\phi_0}T_{\mu\nu}\,,\label{12}
\eeqa
\begin{equation}
\label{13}
  2\Box_{\eta}\psi+\frac{1}{2}\Box_{\eta}{\overline h}+\sigma_{\alpha ,}
  ^{\phantom{\alpha}\alpha}
  +\lambda\phi_0^n[n\phi_0^{-1}-n(n-1)\phi_0^{-2}]\psi+\lambda\phi_0^n[1-n\phi_0^{-1}
  ]=0\,.
\end{equation}
We have discarded the field equation (\ref{5}) since it gives only
higher than linear order terms.

We can eliminate the  term proportional to $\psi_{,\mu\nu}$  by choosing an appropriate
gauge. In fact, by writing the auxiliary field $\sigma_{\alpha}$ as
 \begin{equation}\label{14}
 \sigma_{\mu}=-\psi_{,\,\mu}\,,
 \end{equation}
 Eq.(12) reads
 \begin{equation}\label{15a}
 \Box_{\eta} \overline {h}_{\mu\nu}+\eta_{\mu\nu} \Box_{\eta}
 \psi +\lambda\phi_0^n\left[\eta_{\mu\nu}\left(1+\frac{n}{\phi_0}\psi\right)+
 h_{\mu\nu}\right]\simeq -
 e^{\phi_0}\, T_{\mu\nu}\,.
 \end{equation}
 In our approximations, we can neglect the terms in $h_{\mu\nu}$
 and $\psi$ in Eq.(\ref{15a}) being $h_{\mu\nu}\ll\eta_{\mu\nu}$
 and $\psi\ll 1$. Eq.(\ref{15a}) becomes
 \begin{equation}\label{15}
 \Box_{\eta} \overline {h}_{\mu\nu}+\eta_{\mu\nu} \Box_{\eta}
 \psi +\lambda\phi_0^n\eta_{\mu\nu}\simeq -
 e^{\phi_0}\, T_{\mu\nu}\,.
 \end{equation}

 By defining the further auxiliary field
 \begin{equation}\label{16}
 \tilde {h}_{\mu\nu} \equiv
 {\overline h}_{\mu\nu} +\eta_{\mu\nu} \psi\,,
 \end{equation}
 we get the final form
 \begin{equation}\label{17}
 \Box_{\eta} \tilde{h}_{\mu\nu}+\lambda\phi_0^n\eta_{\mu\nu}\simeq
 -e^{-\phi_0}\, T_{\mu\nu}\,.
 \end{equation}
The original perturbation field $h_{\mu\nu}$ can be written in terms of the new field as

 \begin{equation}\label{18}
 h_{\mu\nu}=\tilde{h}_{\mu\nu}-\frac{1}{2}\,\eta_{\mu\nu} \tilde {h}+
 \eta_{\mu\nu} \psi\,,
 \end{equation}
 being $\tilde{h}\equiv \eta^{\mu\nu} \tilde{h}_{\mu\nu}$.

We turn now to the Klein-Gordon Eq.(\ref{13}). With a little
algebra, it can be recast in the form
\begin{equation}\label{19}
  \left(\Box_{\eta}+c_1^2\right)\psi\simeq
  -\frac{e^{-\phi_0}}{2}T-\Phi_0\,,
\end{equation}
where $T$ is the trace of the stress-energy tensor of standard
matter and the constants are

\beq \label{20}
c_1^2=\lambda\phi_0^{n}\left[n(n-1)\phi_0^{-2}-n\phi_0^{-1}\right]\,,\qquad
\Phi_0=\left(3-n\phi_0^{-1}\right)\lambda\phi_0^{n}\,. \eeq We are
working in the weak-field and slow motion limits, namely we assume
that the matter stress-energy tensor $T_{\mu\nu}$ is dominated by
the mass density term. Furthermore,  we neglect time derivatives
with respect to the space derivatives, so that
$\Box_{\eta}\rightarrow -\Delta$, where $\Delta$ is the ordinary
Laplace operator in flat spacetime. The linearized field equations
(\ref{17}) and (\ref{19}) have, for point--like distribution of
matter\footnote{ To be precise, we can define a Schwarzschild mass
of the form $$M=\int(2T^{0}_{0}-T^{\mu}_{\mu})\sqrt{-g}d^3x \,,$$
and $\rho(r)=M\delta(r)$.} the following solutions
 \begin{equation}\label{21}
 h_{00}\simeq
 -\frac{2G_N M}{r}(1-e^{-c_{1}r})+c_2r^2+c_3\cosh (c_1r)\,,
 \end{equation}
 \begin{equation} \label{22}
 h_{ii} \simeq
 -\frac{2G_N M}{r}(1+e^{-c_{1}r})-c_2r^2-c_3\cosh (c_1r)\,,
 \end{equation}
 \begin{equation}\label{23}
 \psi \simeq
\frac{2G_N M}{r}e^{-c_{1}r}+c_3\cosh (c_1r)\,,
 \end{equation}
 where
 \begin{equation}\label{24}
c_2=2\pi\lambda\phi_0^n\,,\qquad c_3=
\frac{3-n\phi_0^{-1}}{n(n-1)\phi_0^{-2}-n\phi_0^{-1}}\,. \eeq Here
the functions (\ref{21}) and (\ref{22}) are the nonzero components
of $h_{\mu\nu}$ while Eq.(\ref{23}) is the perturbation of the
dilaton. We have restored standard units by using Eq.(\ref{2}).

When $\lambda=0$ and in the slow motion limit,  the
$(0,0)$--component of the field Eq.(\ref{17})
 reduces to the
usual Poisson equation
 \begin{equation}\label{25}
 \Delta\Phi=4\,\pi  G_{N} \rho\,.
 \end{equation}
 Here $\Phi$ is the Newtonian potential which is linked
 to the metric tensor by the relation
$h_{00}=2\,\Phi$.

 What we have obtained are solutions of the linearized field
equations, starting from the action of a scalar field
nonminimally coupled to the geometry, and minimally coupled to
the ordinary matter (specifically the string-dilaton gravity).
Such solutions explicitly depend on the parameters  $\phi_0,
n,\lambda$ which assign the model in the class (\ref{1}).

\section{\normalsize\bf Discussion and Conclusions}

The above solution (\ref{21}), as we said, can be read as a
Newtonian potential with exponential and quadratic corrections \ie
\begin{equation}\label{26}
 \Phi(r)\simeq
 -\frac{G_N M}{r}(1-e^{-c_{1}r})+\frac{c_2}{2}r^2+\frac{c_3}{2}\cosh
 (c_1r)\,.
 \end{equation}

 In general, it can be shown
\cite{stelle},\cite{schmidt},\cite{kenmoku} that most of the
extended theories of gravity has a weak field limit of similar
form, i.e. \beq \label{yukawa} \Phi(r)=-\frac{G_N
M}{r}\left[1+\sum_{k=1}^{n}\alpha_k e^{-r/r_k}\right]\,, \eeq
where $G_{N}$ is the value of the gravitational constant as
measured at infinity, $r_k$ is the interaction length of the
$k$-th component of non-Newtonian corrections. The amplitude
$\alpha_k$ of each component is normalized to the standard
Newtonian term; the sign of $\alpha_k$ tells us if the
corrections are attractive or repulsive (see \cite{will} for
further details). Besides, the variation of the gravitational
coupling is involved. As an example, let us take into account only
the first term of the series in $(\ref{yukawa})$ which is usually
considered the leading term (this choice is not sufficient if
other corrections are needed). We have

 \beq
\label{yukawa1} \Phi(r)=-\frac{G_N M}{r}\left[1+\alpha_1
e^{-r/r_1}\right]\,. \eeq The effect of non-Newtonian term can be
parameterized by $(\alpha_1,\,r_1)$. For large distances, at which
$r\gg r_1$, the exponential term vanishes and the gravitational
coupling is $G_{N}$. If $r\ll r_1$, the exponential becomes unity
and, by differentiating Eq.(\ref{yukawa1}) and comparing with the
gravitational force measured in laboratory, we get

\beq \label{yukawa2}
G_{lab}=G_{N}\left[1+\alpha_1\left(1+\frac{r}{r_1}\right)e^{-r/r_1}\right]
\simeq G_{N}(1+\alpha_1)\,, \eeq where $G_{lab}=6.67\times
10^{-8}$ g$^{-1}$cm$^3$s$^{-2}$ is the usual Newton constant
measured by Cavendish-like experiments. Of course,  $G_{N}$ and
$G_{lab}$ coincide in the standard gravity. It is worthwhile to
note that, asymptotically, the inverse square law holds but the
measured coupling constant differs by a factor $(1+\alpha_1)$. In
general, any  correction introduces a characteristic length that
acts at a certain scale for the self-gravitating systems. The
range of $r_k$ of the $k$th-component of non-Newtonian force can
be identified with the mass $m_k$ of a pseudo-particle whose
Compton's length is

\beq \label{27} r_k=\frac{\hbar}{m_k c}\,. \eeq The interpretation
of this fact is that, in the weak energy limit, fundamental
theories which attempt to unify gravity with the other forces
introduce, in addition to the massless graviton, particles {\it
with mass} which carry the gravitational force \cite{gibbons}.
These masses  introduce length scales which are

\beq\label{28} r_k=2\times 10^{-5}\left(\frac{1\,
\mbox{eV}}{m_k}\right)\mbox{cm}\,. \eeq There have been several
attempts to constrain $r_k$ and $\alpha_k$ (and then $m_k$) by
  experiments on scales in the range $1 \,\mbox{cm}<r< 1000\,
  \mbox{km}$, using
totally different techniques
\cite{fischbach},\cite{speake},\cite{eckhardt1}. The expected
masses for particles which should carry the additional
gravitational force are in the range $10^{-13} \mbox{eV}<m_k<
10^{-5}\, \mbox{eV}$. The general outcome of these experiments,
even retaining only the term $k=1$, is that a ''geophysical
window" between the laboratory and the astronomical scales has to
be taken into account. In fact, the range

\beq |\alpha_1|\sim 10^{-2}\,,\qquad r_1\sim 10^2\div
10^3\,\,\mbox{m}\,,\eeq is not excluded at all in this window. An
interesting suggestion has been given by Fujii \cite{fujii1},
which proposed that the exponential deviation from the Newtonian
standard potential (the ''fifth force") could arise from the
microscopic interaction which couples to nuclear isospin and
baryon number.

The astrophysical counterpart of these non-Newtonian corrections
seemed ruled out till some years ago due to the fact that
experimental tests of general relativity predict ''exactly" the
Newtonian potential in the weak energy limit, ''inside" the Solar
System. Recently, as we said above,
 indications of an anomalous, long--range acceleration
revealed from the data analysis of Pioneer 10/11, Galileo, and
Ulysses spacecrafts (which are now almost outside the Solar
System) makes these Yukawa--like corrections come into play
\cite{anderson}. Besides, Sanders \cite{sanders} reproduced the
flat rotation curves of spiral galaxies by using

\beq\label{sand} \alpha_1=-0.92\,,\qquad r_1\sim
40\,\,\mbox{kpc}\,.\eeq His main hypothesis is that the additional
gravitational interaction is carried by an ultra-soft  boson
whose range of mass is $m_1\sim 10^{-27}\div 10^{-28}$eV. The
action of this boson becomes efficient at galactic scales without
the request of enormous amounts of dark matter to stabilize the
systems.

Eckhardt \cite{eckhardt} uses a combination of two exponential
terms and gives a detailed explanation of the kinematics of
galaxies and galaxy clusters, without dark matter models, using
arguments similar to those of Sanders.

It is worthwhile to note that both the spacecrafts and galactic
rotation curves indications are ''outside" the usual Solar System
boundaries used to test General Relativity. However, the above
authors do not start from any fundamental theory in order to
explain the outcome of Yukawa corrections. In their contexts,
these terms are phenomenological.

Another important remark in this direction deserves the fact that
some authors \cite{mcgaugh} interpret the recent experiments on
cosmic microwave background as BOOMERANG \cite{debernardis} in
the frame of {\it modified Newtonian dynamics} without invoking
any dark matter model.

All these facts point towards the line of thinking that
''corrections" to the standard gravity have to be seriously taken
into account.

Let us turn now to the above solutions (\ref{21})--(\ref{23}), in
particular to the gravitational potential (\ref{26}). This comes
out from the weak field limit (PPN approximation) of a
string-dilaton effective action (\ref{1}). The specific model is
singled out by the number of spatial dimension $d$ and the form of
self-interaction potential $V(\phi)$. We have considered the quite
general class $V(\phi)=\lambda\phi^n$.

Without loosing of generality, we can assume $\phi_0=1$ in
Eq.(\ref{7}). This means that for $\phi=1$ the standard
gravitational coupling is restored in the action (\ref{1}).
However, the condition $\psi\ll 1$ must hold in (\ref{7}). For
the choice $n=3$, we have

\begin{equation}\label{30}
 \Phi(r)\simeq
 -\frac{G_N M}{r}(1-e^{-c_{1}r})+\frac{c_2}{2}r^2\,.
 \end{equation}
 where, beside the standard Newtonian potential, two
 corrections are present. Due to the definition of the constants
 $c_{1,2}$ their strength directly depends on the coupling $\lambda$ of
 the self-interaction potential $V(\phi)$. If $\lambda>0$, from
 Eq.(\ref{20}) and Eq.(\ref{24}), we have that the first
 correction is a {\it repulsive} Yukawa-like term with
 $\alpha_1=-1$ and $r_1=c_1^{-1}=\lambda^{-1/2}$.
 The second correction is given by a sort of
 positive-defined cosmological constant $c_2$ which acts as a repulsive
 force\footnote{We have to note that if in the Poisson equation
 we have a positive
 constant density, it gives rise to a repulsive quadratic potential in $r$
 and then to a linear force. A positive constant density can be easily interpreted as a sort
 of cosmological constant.} proportional to $r$.

 If $\lambda<0$, the first correction is oscillatory while the
 second is attractive.

 From the astrophysical point of view, the first situation is more
 interesting. If we assume that the dilaton is an ultra-soft boson
 which carries the scalar mode of gravitational field, we get, by
Eq.(\ref{28}), that the length scale $\sim 10^{22}\div 10^{23}$
cm, needed to explain the flat rotation curves of spiral galaxy,
is obtained if its mass range is  $m\sim 10^{-27}\div
10^{-28}$eV. The second  correction to the Newtonian potential can
contribute to stabilize the system being repulsive and acting as
a  constant density which is a sort of cosmological constant at
galactic scales (see also \cite{whitehouse} but the models which
they used are different from our). In general, if $\alpha_1\sim
-1$ the flat rotation curves of galaxies can be reconstructed
\cite{sanders}.

If the mass of the dilaton is in the range $10^{-13} \mbox{eV}<m<
10^{-5} \mbox{eV}$ also the ''geophysical windows" could be of
interest. Finally the mass range $m\sim 10^{-9}\div 10^{-10}$eV
could be interesting at Solar System scales (for the allowed mass
windows in cosmology see \cite{gasperini2}). Similar analysis can
be performed also for other values of $n$ which means other
models of the class (\ref{1}).

In conclusion, we have derived the weak energy limit of
string-dilaton  gravity showing that the  Newtonian gravitational
potential is corrected by exponential and quadratic terms. These
terms introduce natural length scales which can be connected to
the mass of the dilaton. If the dilaton is an ultra-soft boson,
we can expect observable effects at astrophysical scales. If it
is more massive, the effects could be interesting at geophysical
or microscopic scales.

\end{document}